\documentclass[12pt,preprint]{aastex}

\usepackage{bm}
\usepackage{xcolor}

\newcommand{\vect}[1]{\boldsymbol{\mathbf{#1}}}
\newcommand{\eb}{\begin{equation}}
\newcommand{\ee}{\end{equation}}

\shorttitle{Double stars in Gaia DR1}
\shortauthors{Makarov et al.}
\begin{document}

\title{Double stars and astrometric uncertainties in Gaia DR1} 
\author{Valeri V. Makarov$^1$, Claus Fabricius$^2$, Julien Frouard$^1$}
\affil{$^1$United States Naval Observatory, 3450 Massachusetts Ave. NW, Washington, DC 20392-5420, USA}
\email{valeri.makarov@usno.navy.mil}
\affil{$^2$Institut de Ci{\'e}ncies del Cosmos, Universitat de Barcelona (IEEC-UB), Mart{\'i} Franqu{\`e}s 1, E-08028 Barcelona, Spain}

\begin{abstract}
A significant number of double stars with separations up to
2.5 arcsec are present in the Gaia Data Release 1 astrometric catalogs. Limiting our analysis to a well-studied sample
of 1124 doubles resolved by Hipparcos, provided with individual Tycho component photometry, and cross-matched with the
TGAS catalog, we estimate a rate of at least 3\% for brighter double stars in Gaia DR1, which should be resolved
in the future data releases. Gaia astrometric results are affected by unresolved
duplicity. The variance-normalized quadratic differences of proper motion between Gaia and Hipparcos do not follow the expected $\chi^2$
distribution and show signs of powerful degradation in the components aligned with the axes of the double systems. This concerns only
pairs with separation below 1.2 -- 1.5 arcsec, which mostly remain unresolved in Gaia DR1. On the other hand, the orthogonal
proper motion components and parallaxes do not have any detectable perturbation, as well as all astrometry for separations
above 1.5 arcsec. Gaia parallaxes do not seem to be perturbed by duplicity, with Gaia - Hipparcos differences being systematically smaller than the expectation. The rate of incorrectly identified, or swapped, companions is estimated at 0.4\%.
\end{abstract}


\section{Introduction}
\label{Introduction}
The first release of the Gaia mission data (Gaia DR1) includes accurate positions, parallaxes, and proper motions for more than 2 million
Tycho-2 and Hipparcos stars and positions only for the larger Gaia sample of 1.1 billion objects \citep{bro}. The former part, called
TGAS, was constructed differently from the large catalog \citep{lin}. The proper motions of TGAS stars in DR1 were computed using
Hipparcos and Tycho-2 positions at, or close to, the epoch J1991.25, and Gaia own position determinations around J2015. Thus, the
TGAS astrometric solution is only partially independent of the previous Hipparcos and Tycho-2 catalogs. However, Gaia DR1 proper
motions and parallaxes are presumed to be practically independent of their counterparts in the Hipparcos catalog, because the latter were not
explicitly used in the astrometric reductions\footnote{A small correlation of Hipparcos and Gaia proper motions may still be present
because of the nonzero covariances of positions and proper motions in each catalog.}. A conscious effort was made at the catalog
production stage to improve the reliability of DR1 astrometry in TGAS (at the expense of completeness) by filtering out all sources with
standard formal errors of parallax in excess of 1 mas or 20 mas in positions, per coordinate. It is important to note,
in the context of this paper, that this filtering was not meant to discard specifically blended double sources or other perturbed
images, because the formal errors are not sensitive to the observed scatter of residuals.
As explained by \citet{fab}, some double star and optical pair components could be removed as parts of duplicated entries mostly
originating from redundant entries in the Initial Gaia Source List. The threshold near-neighbor distance for this additional filtering 
(typically, 59 mas) was
lower than the separation of double stars resolved by Hipparcos and Tycho. Therefore, it is not surprising that some known
double stars with separations below $2\farcs5$
survived this filtering and were included in the DR1, and sometimes, with both components as separate entries.

The goal of this paper is to estimate the frequency of unresolved double stars in Gaia DR1 catalogs using a well defined and reliable
sample of pairs previously resolved by Hipparcos and Tycho, and to investigate the expected astrometric degradation caused by
unresolved duplicity. A similar validation effort is briefly discussed by \citet{are}. They note occasional gross errors in astrometry
and photometry resulting from Gaia sources being cross-identified with the wrong components of resolved pairs. Here, we compare
Gaia DR1 data with the original edition of the Hipparcos catalogue \citep{esa}
because the later re-processing by \citet{van}, which was
used by \citet{are}, contains strongly underestimated standard errors for brighter stars \citep{fro,are}. This inflates the $\chi^2$ statistics of astrometric differences
in an external comparison and hides the additional perturbation caused by duplicity.

\section{The sample}
\label{sample.sec}
We start with the relatively well-studied collection of $9\,473$ components of double and multiple systems resolved by Hipparcos and listed in
\citet{fm}. The Double and Multiple Systems Annex (DMSA) of the main Hipparcos catalog lists over $12\,000$ double and multiple systems
with separations between 0.1 and 2.5 arcsec. More than 10\% of Hipparcos stars were resolved with these separations. The special
photometric solution for these systems, which was a derivative of the Tycho-2 project \citep{hoga, hogb}, produced 
separate $B_T$ and $V_T$ magnitudes \citep[similar to Johnson $B$, $V$, see][]{bes}
for components of more than $7\,000$ systems, but only $5\,173$ systems with separations greater than 0.3 arcsec were ever published.
We choose this well-defined and reliable sample over the larger DMSA collection because the latter contains a fraction of low-quality
and suspicious solutions at smaller angular separations, and because the additional color information allows one to estimate
more subtle photocenter effects for unresolved doubles.

Each of the $9\,473$ components in the initial sample was cross-matched with TGAS using HIP identification numbers. 
The number of components found in TGAS is $2\,768$. Thus, approximately 29\% of the resolved double and multiple
systems with separations between $0\farcs3$ and $2\farcs5$ are present in TGAS. Using the rate of resolved doubles in Hipparcos and this
estimate, we should expect roughly 3\% double or multiple stars to be present in Gaia DR1 at brighter magnitudes ($G\leq 13$). 
At fainter magnitudes, the rate of optical pairs goes up and field confusion becomes a significant factor. Since the DR1 pipeline was not set up to
specially treat close double images, a significant fraction of the astrometric catalog may be impacted.

Only one TGAS entry was found for each cross-matched double in the sample\footnote{In fact, each unique HIP number appears only
once in TGAS.}. This implies that DMSA systems having the same HIP number are never resolved in TGAS.
In most cases, the matched TGAS entry corresponds to the brighter component, but sometimes the magnitudes are close and the wrong
component is identified in TGAS. The secondary Gaia DR1 catalog (with positions only) does include sources matching some of the secondary
components. By inspecting a random sample, we found that most of the doubles with separation above $1\farcs5$ were resolved in the secondary
catalog, while most of those with smaller separations were not. Thus, an effective threshold resolution of the Gaia DR1 catalog is
about $1\farcs5$ at brighter magnitudes, although much closer pairs are sometimes resolved into separate entries too. The separation limit may be
higher for the fainter general DR1 stars. For a discussion of the general angular resolution of Gaia DR1, see
also Sect 4.4 and Fig. 17 of \citet{are}. In the following, we will concentrate on
the $2\,768$ cross-matched TGAS components in an attempt to detect the impact of duplicity on the astrometric performance.

\begin{figure}[htbp]
  \centering
  \includegraphics[angle=0,width=0.7\textwidth]{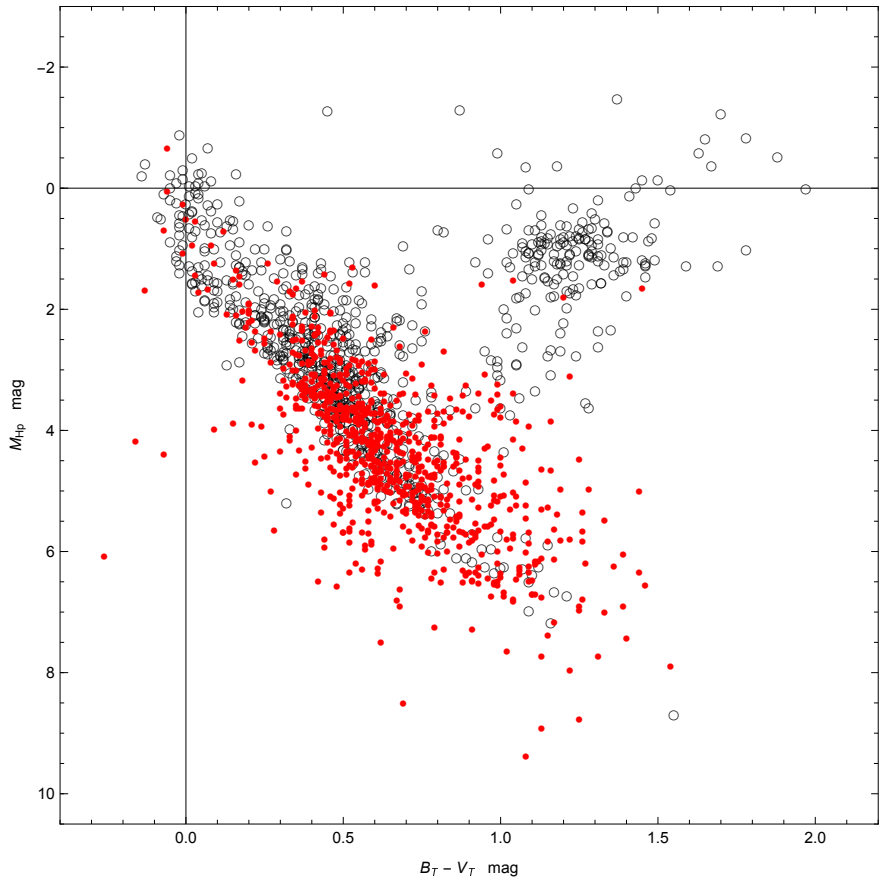}
\caption{HR diagram of 1676 components of double stars with Tycho $BV$ magnitudes and Gaia DR1 parallaxes. Primary components are shown with
open circles, secondary with red dots. \label{hr.fig}}
\end{figure}

As an intermediary verification step, we can use the TGAS parallaxes (much superior to Hipparcos parallaxes), Hipparcos $Hp$ magnitudes, and the
Tycho-2 $B_T$ and $V_T$ magnitudes for individual components, to construct an improved HR diagram for close double stars.
Fig.~\ref{hr.fig} shows a diagram for 838 pairs with statistically significant parallaxes of primaries, $\varpi/\sigma_\varpi >7$. The
primary (A) components are marked with open circles, and the secondaries with red dots. The parallax of a secondary is always assumed to be equal
to the parallax of the primary found in TGAS. This is correct for physical binaries but is wrong for optical pairs, which should be
dispersed across the diagram. Indeed, as the primaries show a well-defined main sequence, red clump, and a giant branch, the secondaries
seem to be more dispersed around the main sequence. Some of the deviant secondaries may be optical pairs,
others may reflect occasional gross errors in the $B_T$, $V_T$ magnitudes.
As a note in passing, several supergiants are present among the primaries, and a few
possible hot subdwarfs (sdB) among the secondaries. Some of these special cases deserve a separate study, but the general conclusion
is that most of the systems in the sample are physical binaries and the quality of Tycho photometry and Gaia astrometry is high. This result also
confirms that the majority of resolved double systems are matched with correct TGAS sources.

\begin{figure}[htbp]
  \centering
  \plotone{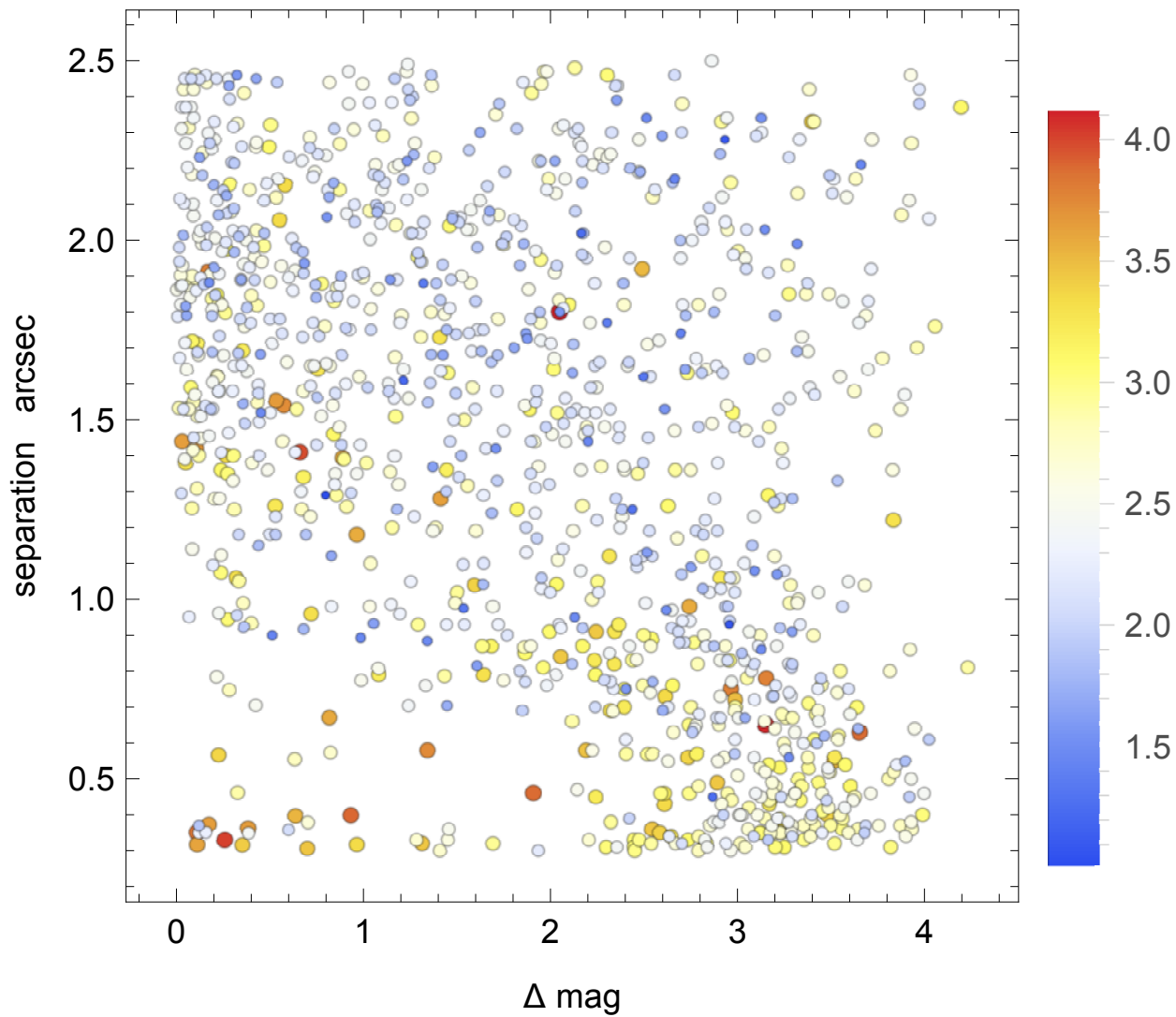}
\caption{Distribution of separations and $H_p$ magnitude differences between the components of 1124 double systems resolved in Hipparcos
and present as single entries in TGAS. The color of each bubble represents the decimal logarithm of the ``excess noise significance" parameter 
given in the
Gaia catalog. The temperature map scheme and the range of parameters are shown in the color bar legend. \label{sep.fig}}
\end{figure}

Some of the DMSA systems are represented with only one component in the Tycho-2 sample. Such lonely components were included when
the secondary was too faint for a confident detection in the Tycho observations, i.e., when the signal-to-noise ratio was below
a threshold level. The number of complete Tycho pairs in the set matched with TGAS is 1124, but many of them have not been
resolved by Gaia. What are the separations of these pairs?
Fig. \ref{sep.fig} shows the distribution of these doubles in the $\Delta H_p$ magnitude versus separation plane. Surprisingly, we do not see
a pronounced bias of the doubles with at least one matching entry in Gaia DR1 toward larger $\Delta$mag or smaller separation, although a small pile-up is present
in the lower right corner of the graph. This confirms that the decision whether to include a star in Gaia DR1 was not directly
related to the shape of the image. On the other hand, pairs with small magnitude difference may have a better chance to make it to the release, 
because they need the same exposure time (or gate setting). Perhaps, the ``significance of excess noise"
parameter $D$ given in TGAS should be more informative about the degree of perturbations observed for a given object.
If $D>2$, the amount of extra noise is statistically significant, and only a few percent of regular, unperturbed stars are
expected to have it. The range of $D$ is between 10 and $13\,000$ for our sample indicating a high degree of perturbation and extra
noise. The color of the bubbles in Fig. \ref{sep.fig} represents the $D$ value in a temperature map, starting with the blue for
the smallest values and ending with the deep red for the largest. We observe some tendency of most extreme extra noise parameters to occur
at small separations. 

\section{Astrometric errors}

As explained in \citet{lin,mig,mak}, the covariances of bivariate parameters should be used when comparing data in two independent
catalogs and estimating the statistical significance of the differences. In our case, the positions and proper motions in both the Hipparcos
and Gaia catalogs have complete covariance matrices, including nonzero correlations. For example, if we want to estimate the statistical
significance of a proper motion difference vector $\Delta_{\vect{p}} = \vect{p}_{\rm Gaia} - \vect{p}_{\rm HIP}$, where $\vect{p}_{\rm HIP}=
[\mu_{\alpha *}, \mu_\delta]_{\rm HIP}$ is the proper motion vector composed of the tangential components in right ascension and declination, provided
in the Hipparcos catalog (and likewise for Gaia), the variance-normalized quadratic difference is computed as
\eb
u_{\vect{p}}=\Delta_{\vect{p}}\, \vect{C}^{-1} \, \Delta_{\vect{p}}^T,
\label{u.eq}
\ee
where $\vect{C}$ is the total covariance matrix of the difference vector, $\vect{C}=\vect{C}_{\rm Gaia}+\vect{C}_{\rm HIP}$. 
The quadratic form $u$ is $\chi^2$-distributed with 2 degrees of freedom if the
measurements are normally distributed. Normalized differences of positions can be computed in a similar way, making sure that
the correct 2$\times2$ blocks of the $5\times5$ formal covariance matrices are taken. For each star, the $u$ statistic is a random number.
The sample distribution of a set of $u$ values allows us to assess how closely the formal errors estimate the actual uncertainty of the
data. Since parallaxes are univariate statistics, their variance-normalized differences are computed simply by

\eb
u_\varpi=\frac{(\varpi_{\rm Gaia}-\varpi_{\rm HIP})^2}{\sigma_{\varpi, {\rm Gaia}}^2+\sigma_{\varpi, {\rm HIP}}^2},
\label{par.eq}
\ee
and the sample distribution is expected to be close to the $\chi^2$ distribution with one degree of freedom.

\begin{figure}[htbp]
  \centering
  \plottwo{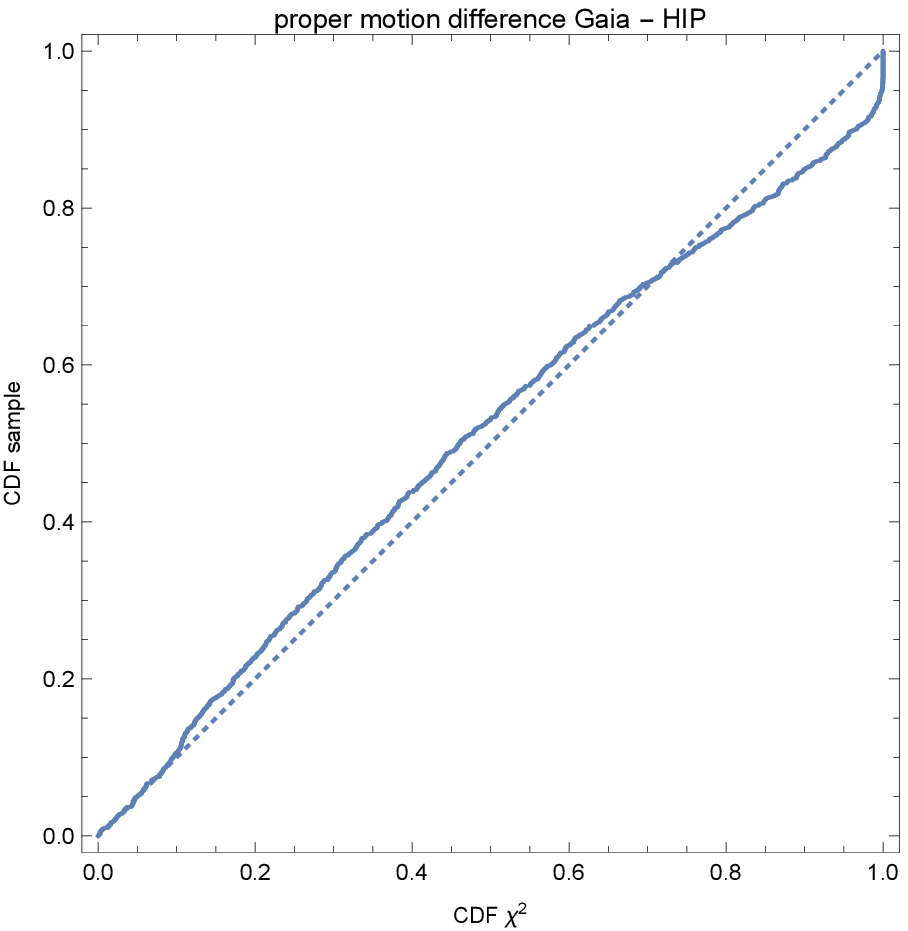}{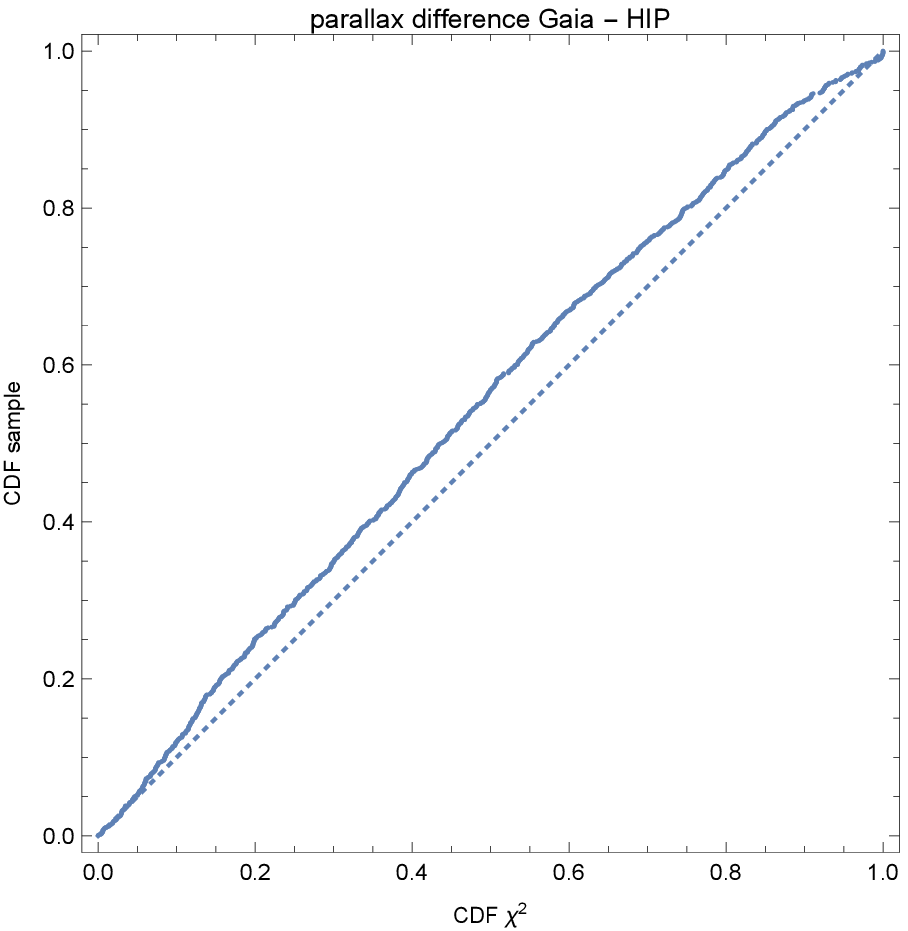}
\caption{Probability plots, i.e., sample CDF versus expected CDF, for normalized variances ($\chi^2$) of Gaia$-$Hipparcos differences
in proper motions (left) and parallaxes (right) for 1124 double stars. \label{cdf.fig}}
\end{figure}

Fig. \ref{cdf.fig} shows the probability plots of the $u$ statistics for the proper motion (left) and parallax (right) differences
between TGAS and Hipparcos for the 1124 components of double stars resolved by Hipparcos. A probability plot is a way of graphically
representing any differences between a sample distribution and an expected theoretical distribution, where the sample cumulative distribution
function  (CDF) is mapped versus the theoretical CDF. For a perfect match of the distributions, the ordered sample quantiles should lie
on the diagonal which is marked with a dashed line. We find in both cases a definite departure from the expectancy. For all parallaxes,
and for 70\% of the smaller normalized differences in proper motions, the curves buckle up. This means that the observed scatter of the
differences is {\it smaller} than the formal errors suggest. In other words, all parallax errors, and many of the proper motion errors, are 
{\it overestimated}. This result is puzzling, as formal errors are usually underestimated in astrometric catalogs because it is
difficult to take into account systematic and correlated errors. The largest 30\% of proper motion differences, on the other hand, are
too large compared with the expected probability. This is what we should expect if the astrometric data is additionally perturbed by
the duplicity, which frequently remained unresolved in Gaia DR1.

Formal standard errors were inflated a posteriori in TGAS in an attempt to account for sources of uncertainty apart from the photon
statistics \citep{lin}. The inflation factor (greater than 1.4) is a function of only the parallax formal error, therefore, this
manipulation can be reversed. To clarify our results for this external comparison, we deflated the parallax errors and re-constructed the
probability plots. The resulting curve is slightly closer to the diagonal compared to Fig. \ref{cdf.fig}, right, but the sample distribution
is still confidently deviant from CDF$_{\chi^2[1]}$. This is because the formal errors in TGAS are typically smaller than the errors in Hipparcos,
and scaling the former down in the denominator of Eq. \ref{par.eq} does not help much. We further extended our comparison to the general
TGAS-HIP sample. A similar probability plot for $78\,755$ bona fide single stars (without any signs of binarity or other known perturbations),
not reproduced in this paper, is the opposite to Fig. \ref{cdf.fig}, right, in that the curve is systematically below the diagonal for the entire
interval of cumulative probability. The conclusion is inescapable: the formal errors of parallaxes and most proper motions are
overestimated for double star components, but underestimated for bona fide single stars. The only credible explanation we can find is
that the Hipparcos formal errors were artificially inflated for double stars in the special component solution, and apparently, too much.

This makes finding any additional noise in TGAS data rather difficult. The position differences are practically useless because we would have
to compare positions separated by 23.75 years and bridged by the proper motion, which was derived using the same positions. This
inconvenience opens up another opportunity though. We expect the TGAS positions of unresolved pairs to be displaced toward the photocenters
located on the lines connecting the components. The amount of displacement depends on the $\Delta$mag in the $G$ passband (in which
the astrometric observations were taken) and the separation.
They can be large running up a fraction of $1\arcsec$. These photocenter shifts should perturb the TGAS proper motions, but only in the
directions of double systems, by up to $\sim25$ mas yr$^{-1}$, which should be easily detectable in the Gaia-HIP differences. Thus, our task
now is to compute the normalized proper motion differences in the $s$- (along the lines connecting the components) and the $c$-direction
(orthogonal) for each pair.

This task is performed as follows. Let $\theta$ be the position angle of the double as specified in DMSA. The $s-$direction is then
defined as the unit vector $\vect{s}=[\sin\theta,\, \cos\theta]$, and the orthogonal direction is $\vect{c}=[\cos\theta,\, -\sin\theta]$.
The sought normalized difference components $u_{\vect{s}}$ and $u_{\vect{c}}$ are computed by Eq. \ref{u.eq} replacing the proper motion
difference vector $\Delta_{\vect{p}}$ with its projections $(\Delta_{\vect{p}}\cdot \vect{s})\,\vect{s}$ and $(\Delta_{\vect{p}}\cdot \vect{c})\,
\vect{c}$, respectively. These statistics are expected to be distributed as $\chi^2$ with one degree of freedom.
\begin{figure}[htbp]
  \centering
  \plotone{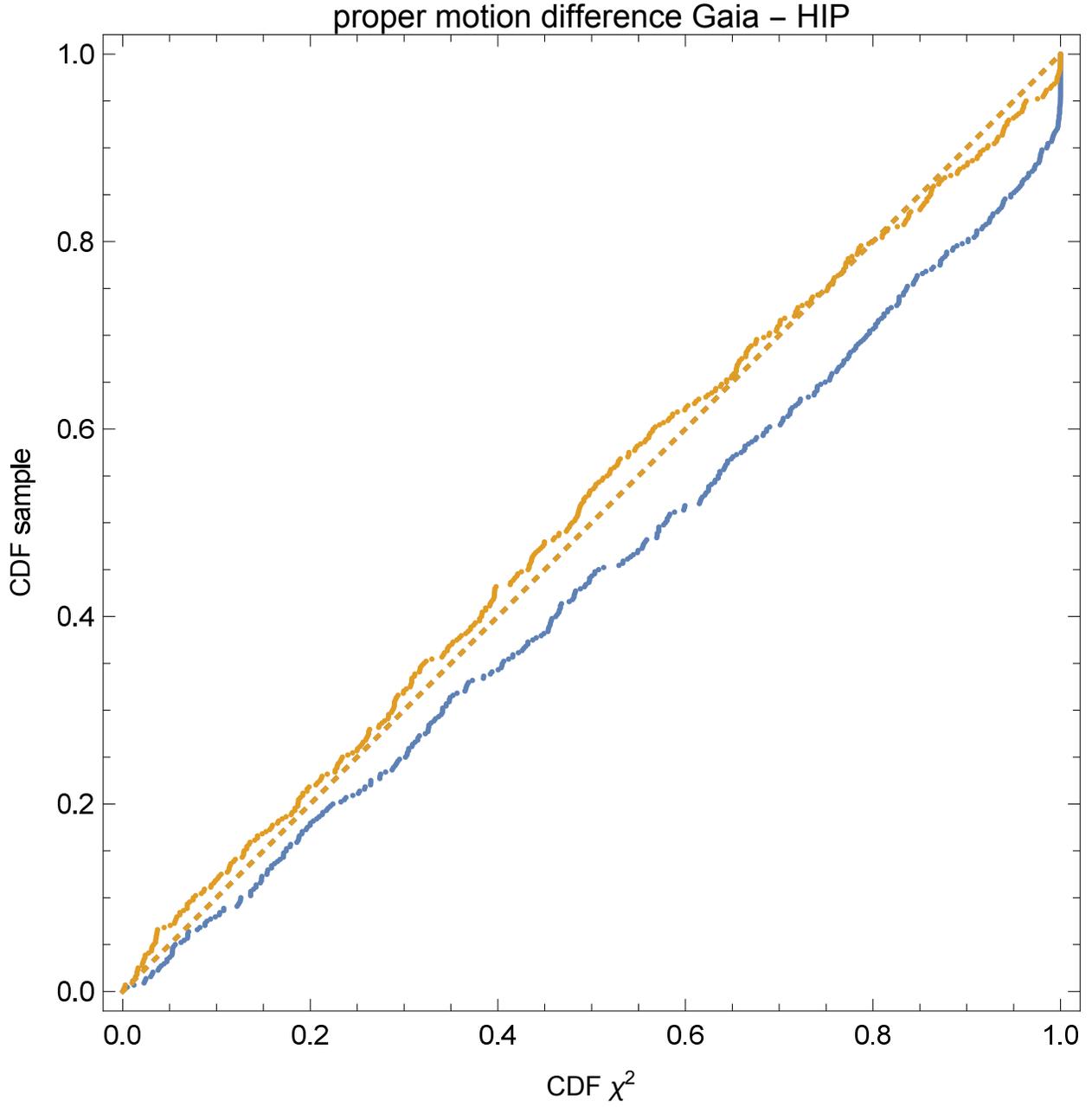}
\caption{Probability plot for normalized variances ($\chi^2$) of Gaia$-$Hipparcos differences
in the $s$-components (blue line) and $c$-components (yellow line) of proper motions . \label{com.fig}}
\end{figure}

Fig. \ref{com.fig} shows the resulting probability plot for $u_{\vect{s}}$ (blue curve) and $u_{\vect{c}}$ (golden curve) for all
pairs with separations less than $1\farcs2$, which are mostly unresolved in Gaia DR1. This tighter threshold is chosen to
emphasize the difference between the components, which is more obvious for smaller separations, as we will see
in the following. While the orthogonal components of proper
motion differences show little perturbation and are distributed similarly to the overall sample, the aligned components display
a very strong additional scatter. To make sure that it is not a mere trick of the eye, we computed 6 different hypothesis tests on the sample distributions
comparing them with the $\chi^2[1]$. The $p-$values from all 6 tests are very small for the $s-$components, with the largest value 
0.0016 produced by the Kuiper's test. On the contrary, the $p-$values for the $c-$components are significant, with a $p=0.61$
from the Pearson's $\chi^2$ and $p=0.55$ from the Kolmogorov-Smirnov test. This obvious difference in dispersion of components
disappears if we reproduce this analysis for all pairs with separations greater than $1\farcs5$. In fact, the $s-$components
are slightly less dispersed than the $c-$components. Thus, no detectable perturbation in Gaia DR1 astrometry emerges for wider
double stars, which are mostly resolved.

\begin{figure}[htbp]
  \centering
  \plotone{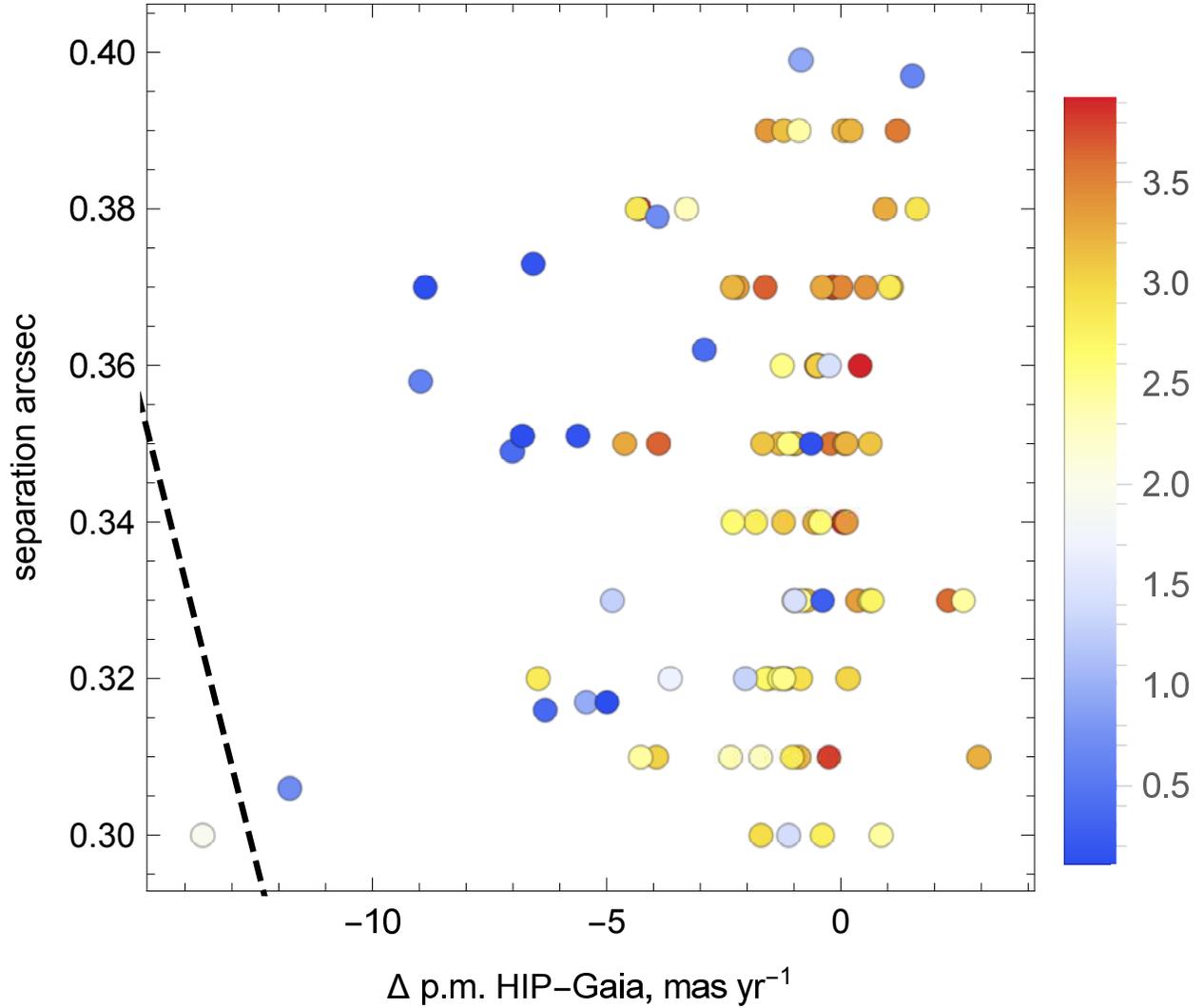}
\caption{Distribution of separations and $s$-components of HIP$-$Gaia proper motion differences for the closest double stars
resolved in the TychoBV sample. The color of each bubble represents the $Hp$ magnitude difference between the star components
ranging from 0 mag (blue) to 4 mag (red), as per the color bar legend. The dashed line shows the expected loci of swapped
components in TGAS. \label{sma.fig}}
\end{figure}

Since the photocenters of unresolved double stars in TGAS are shifted from the primaries toward the secondaries, the errors in the
$s$-components of TGAS proper motions should be predominantly positive. This is nicely confirmed in Fig. \ref{sma.fig} where
the HIP$-$Gaia $s$-component differences are displayed for the smallest separations available in the TychoBV sample. The effect
of unresolved duplicity dramatically increases towards the smallest separations, which is a predictable consequence of the
implemented scheme of PSF centroiding. Despite the random errors somewhat blurring the picture, the great majority of differences
are negative at separations below $0\farcs4$, and the error can reach $15$ mas yr$^{-1}$ in absolute units. Pairs of small magnitude difference
(blue bubbles) tend to produce larger proper motion errors in TGAS, as expected. The dashed line shows the expected proper motion
difference for TGAS entries that were mistakenly cross-matched with the secondary components instead of the primaries (swapped components),
i.e., the separation divided by 23.75 with the negative sign. Two stars in this range of separation are close to this line and are likely to be such
swapped pairs.
\section{Conclusions}
\label{conc.sec}
We have detected, with practically absolute confidence, a strong perturbation of Gaia DR1 positions along the lines connecting
components of unresolved doubles, which resulted in corrupted proper motion components in the corresponding directions. From the
general rate of resolved double and multiple stars in Hipparcos (10\%) and the rate of doubles in the TGAS solution, we estimate
that 3\% or more of the entries in Gaia DR1 should be affected by unresolved duplicity. The parallaxes, on the other hand, do not
show any degradation in accuracy, although this resilience could be explained to some extent by artificially inflated formal
errors in the Hipparcos component solution. We also find that $1\farcs5$ is roughly the threshold separating mostly resolved
doubles from mostly unresolved at brighter magnitudes, although this boundary is rather fuzzy. When a pair of stars is resolved (with the primary
in TGAS and the secondary in the main catalog), the astrometric parameters of the primary seem to be unaffected within the
uncertainties of the two catalogs.

Swapped or misidentified companions represent
another source of crude errors in Gaia DR1. The rate of these can be estimated using the $s-$components of proper motion
differences in Hipparcos
and Gaia, which should be close to the separation (as determined by Hipparcos) divided by the epoch difference, 23.75 yr.
Selecting the differences within $\pm3$ mas yr$^{-1}$ of that value for the entire collection of 1124 pairs,
we find 4 credible cases of such misidentified components, namely, HIP 23317, 50305, 67186, and 114504.
The largest separation, and, hence, the proper motion error is found for the HIP 50305 ($1\farcs409$). The estimated rate of such 
misidentifications is approximately 0.4\% for the collection of Hipparcos resolved doubles.

\acknowledgments
This work has made use of data from the European Space Agency (ESA)
mission {\it Gaia} (\url{http://www.cosmos.esa.int/gaia}), processed by
the {\it Gaia} Data Processing and Analysis Consortium (DPAC,
\url{http://www.cosmos.esa.int/web/gaia/dpac/consortium}). Funding
for the DPAC has been provided by national institutions, in particular
the institutions participating in the {\it Gaia} Multilateral Agreement. We made use of the Department of Defense Celestial 
Database of the USNO Astrometry Department. This research has made use of the VizieR catalogue access tool, CDS,
 Strasbourg, France. The original description of the VizieR service was
 published in A\&AS 143, 23. This work was supported by the MINECO (Spanish Ministry of Economy) -
FEDER through grant ESP2016-80079-C2-1-R and ESP2014-55996-C2-1-R and
MDM-2014-0369 of ICCUB (Unidad de Excelencia 'Mar\'ia de Maeztu'). Discussions with A. Tokovinin helped to improve
the quality of this paper.

\end{document}